\documentstyle[sprocl]{article}

\input{psfig.sty}

\def\Journal#1#2#3#4{{#1} {\bf #2}, #3 (#4)}

\def\PRL{\em Phys. Rev. Lett.}

\def\EPJC{{\em Eur. Phys. J.} C}

\def\be{\begin{equation}}
\def\ee{\end{equation}}
\def\bea{\begin{eqnarray}}
\def\eea{\end{eqnarray}}
\def\fs{\footnotesize}

\begin{document}
\vspace*{-3.cm}
\begin{raggedleft}
BONN-HE-2000-01\\
\hspace*{9.cm}June 2000
\end{raggedleft}
\vspace*{2.0cm}
\title{LEPTON FLAVOUR VIOLATION AT HERA}

\author{R. KERGER\footnote{Supported by a grant from the
German ``Bundesministerium f\"ur Bildung und Forschung''.} \\
(on behalf of the ZEUS and H1 collaborations)} 

\address{Physikalisches Institut der Universit\"at Bonn, Nu{\ss}allee 12,
D-53115 Bonn, Germany\\E-mail: kerger@physik.uni-bonn.de} 


\maketitle\abstracts{The $e^+p$-data collected with the ZEUS
(${\cal L}=47.7$\,pb$^{-1}$) and the H1 (${\cal L}=37$\,pb$^{-1}$)
detectors at HERA in 1994--1997 are analysed for signals of lepton
flavour violation mediated by leptoquark exchange, both in the muon
and the tau channels. No evidence for lepton flavour violation is
found and limits on the leptoquarks' Yukawa couplings are set.}

\vspace*{-1.0cm}\section{Introduction}
Within the standard model (SM) all interactions conserve
lepton flavour individually, allowing us to assign the leptons to
three distinct generations. It is, however, not clear if lepton
flavour is a fundamental quantum number since it has not yet been
brought into relation with an underlying symmetry. Many extensions of
the SM therefore contain lepton flavour violation. Recently,
Super-Kamiokande \cite{sk} has reported evidence for oscillation of
atmospheric neutrinos. This is the first experimental observation of
lepton flavour violation (LFV).\\ 
At HERA, LFV could occur in $eq_1\to \ell q_2$ scattering, the typical
signature being an isolated higher-generation lepton ($\ell=\mu,
\tau$) instead of the scattered electron. This process could be
mediated by leptoquarks with  mass $M_{LQ}$, which allow couplings
both to $(e q_1)$ and to $(\ell q_2)$ pairs. If $M_{LQ}<\sqrt{s}$ 
($\sqrt{s}\approx 300$\,GeV being the HERA centre-of-mass energy), LQs are
predominantly produced in the $s$-channel. In this case, using the
narrow-width approximation (NWA), the resonant production cross
section, $\sigma_{\rm{NWA}}$, is proportional to
$\lambda_{eq_1}^2\times {\mathrm{BR}}_{\ell q_2} \times
q(x=\frac{M_{LQ}^2}{s})$, $\lambda_{eq_1}$ being the Yukawa coupling
at the LQ production vertex, BR$_{\ell q_2}$ the branching
ratio for the decay $LQ \to \ell q_2$ and $q$ the quark density. If
$M_{LQ}\gg\sqrt{s}$, both the $s$- and $u$-channels contribute to the
cross section. Since the propagator contracts to an effective
four-fermion interaction, the cross section is proportional to
$\left[\frac{\lambda_{e q_1}\lambda_{\ell
q_2}}{M_{LQ}^2}\right]^2\equiv [\Psi_{q_1q_2}^{\ell}]^2$.
\vspace*{-0.4cm}\section{Analysis and Results}  
\subsection{$e \leftrightarrow \mu$}   
The main signature consists of a high-transverse-momentum muon
together with a jet and the absence of an isolated electron. After the
preselection, H1 finds 4 ($\mu$+jet)-events, compared to the
SM-expectation of $0.6\pm 0.1$ events. These events \cite{high_pt}
are, however, not consistent \cite{h1_lfv} with a final state as
expected for LQ processes and are removed by the final selection cuts
(requiring among other things $\mu$ and the hadronic final state to be back to back
in azimuth). Both H1 and ZEUS finally observe no candidate.\\ 
\begin{minipage}[t]{6.cm}
\raisebox{-53.mm}{\psfig{figure=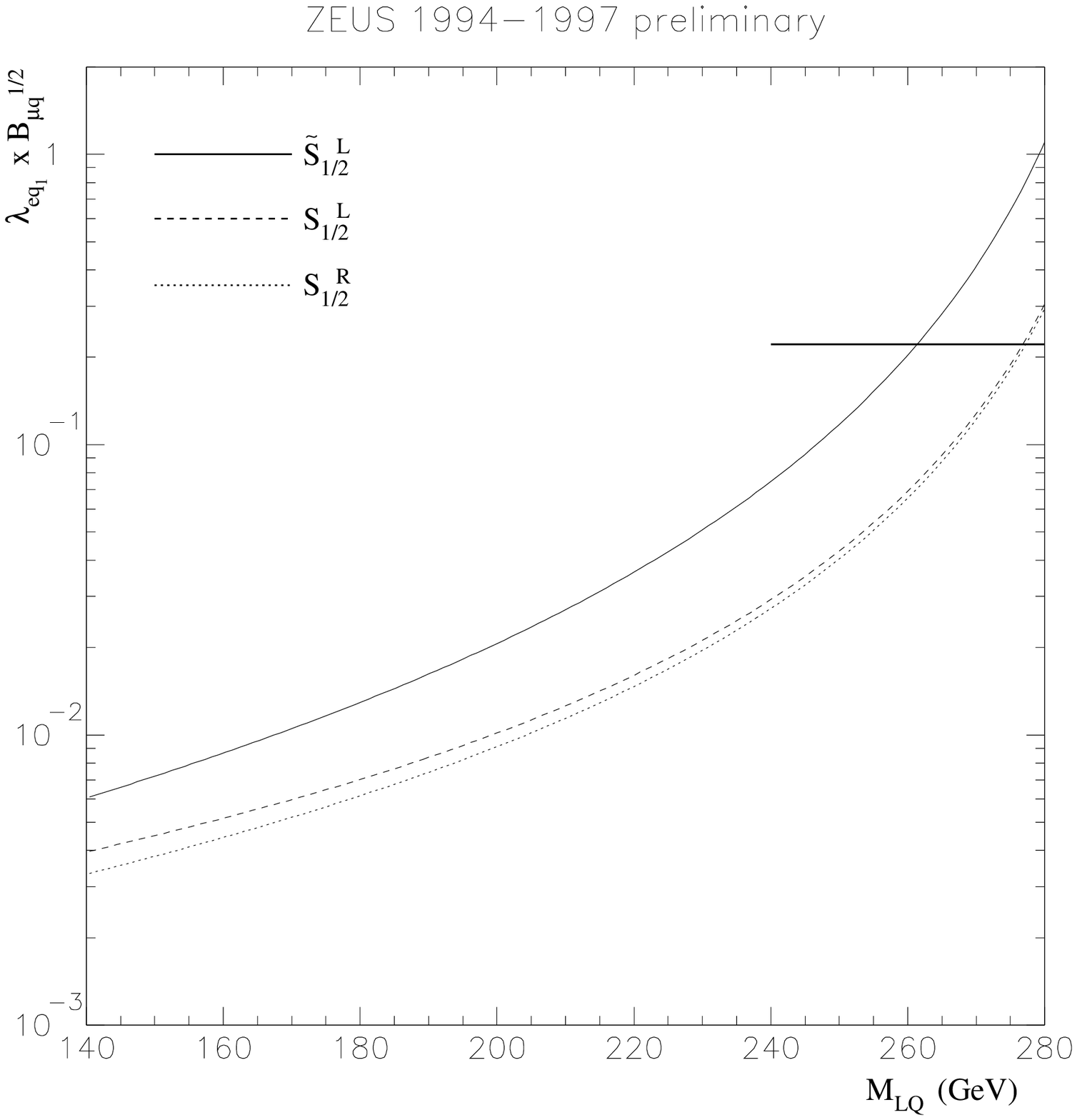,height=6.cm}}
\begin{minipage}[t]{5.5cm}
\footnotesize{Figure 1: 95 $\%$ CL upper limit on $\lambda_{eq_1}\times 
\sqrt{{\mathrm BR}_{\mu q_2}}$ for scalar leptoquarks with fermion
number F=0. $\tilde{S}_{1/2}^L$ couple to $d$-quarks only,
${S}_{1/2}^L$ couple to $u$-quarks only and ${S}_{1/2}^R$ couple to
both $u$- and $d$-quarks.}
\end{minipage}
\end{minipage}\hfil
\begin{minipage}[t]{5.5cm}
\noindent Figure 1 \cite{zeus_lfv} displays the limits on
$\lambda_{e q_1}\times \sqrt{{\mathrm BR}_{\mu q_2}}$ for different
resonantly produced F=0 scalar leptoquarks as a function of the LQ
mass; the areas above the lines are excluded at 95$\%$ confidence
level (CL). If electromagnetic coupling strength
$\lambda_{eq_1}=\sqrt{4\pi\alpha}$ and ${\mathrm BR}_{\mu q_2}$=0.5
are assumed (as indicated by the horizontal bar in Fig.\ 1), masses of
leptoquarks up to 260 -- 280\, GeV, depending on the LQ type, are
excluded. This complements the results obtained by the TeVatron
experiments \cite{d0_mu} which exclude flavour-diagonal LQs decaying
only into $\mu + q$ up to masses of 200\,GeV. 
\end{minipage}
\vspace*{-3.mm}\subsection{$e \leftrightarrow \tau$}
The searches in the $\tau$-channel have to use the final-state
properties of the $\tau$ since its decay vertex cannot be
resolved. Hadronic $\tau$ decays are identified by requiring a narrow
collimated jet with 1 to 3 reconstructed tracks. Furthermore, events
of that kind are required to have a net transverse momentum aligned in
azimuth with the narrow jet associated with the hadronic $\tau$ decay
products. Leptonic $\tau$ decays are identified by requiring, in
addition to a large missing $p_t$, a high-$p_t$ charged lepton in the
missing-$p_t$ direction. ZEUS takes all decay modes into account; H1
considers the hadronic decays separately -- the muonic $\tau$ decays are
covered by the $e \leftrightarrow \mu$ analysis, whereas the $\tau \to
e \nu \bar{\nu}$ decays are not used. No event survives the selection.
For several vector LQs, Fig. 2 displays
the mass-dependent upper limits on $\lambda_{e q_1}\times
\sqrt{{\mathrm BR}_{\tau q_2}}$.  By assuming a Yukawa coupling of 
electromagnetic strength (as indicated by the horizontal bar in the
plot) and ${\mathrm BR}_{\tau q_2}=0.5$, LFV LQs with masses
smaller than 265--285 GeV, depending on the type, can be excluded. \\
\begin{minipage}[t]{5.7cm}
\psfig{figure=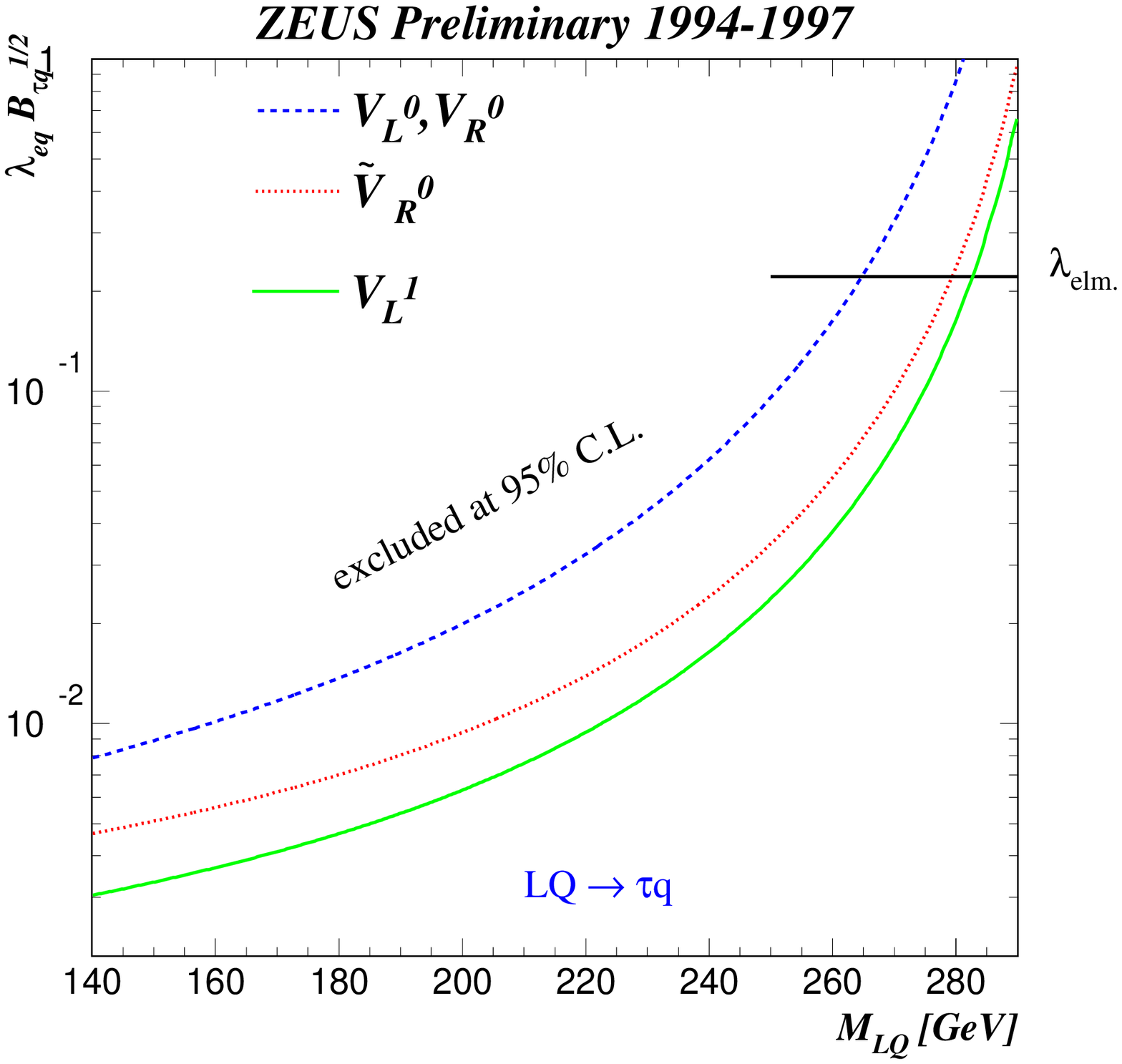,height=5.cm}
\begin{minipage}[t]{5.2cm}
{\fs Figure 2: 95$\%$ CL ZEUS upper limits on
$\lambda_{eq_1}\times\sqrt{{\mathrm BR}_{\tau q_2}}$ for F=0 vector LQ
as a function of LQ mass.}
\end{minipage}
\end{minipage}\hfil
\begin{minipage}[t]{5.7cm}

\psfig{figure=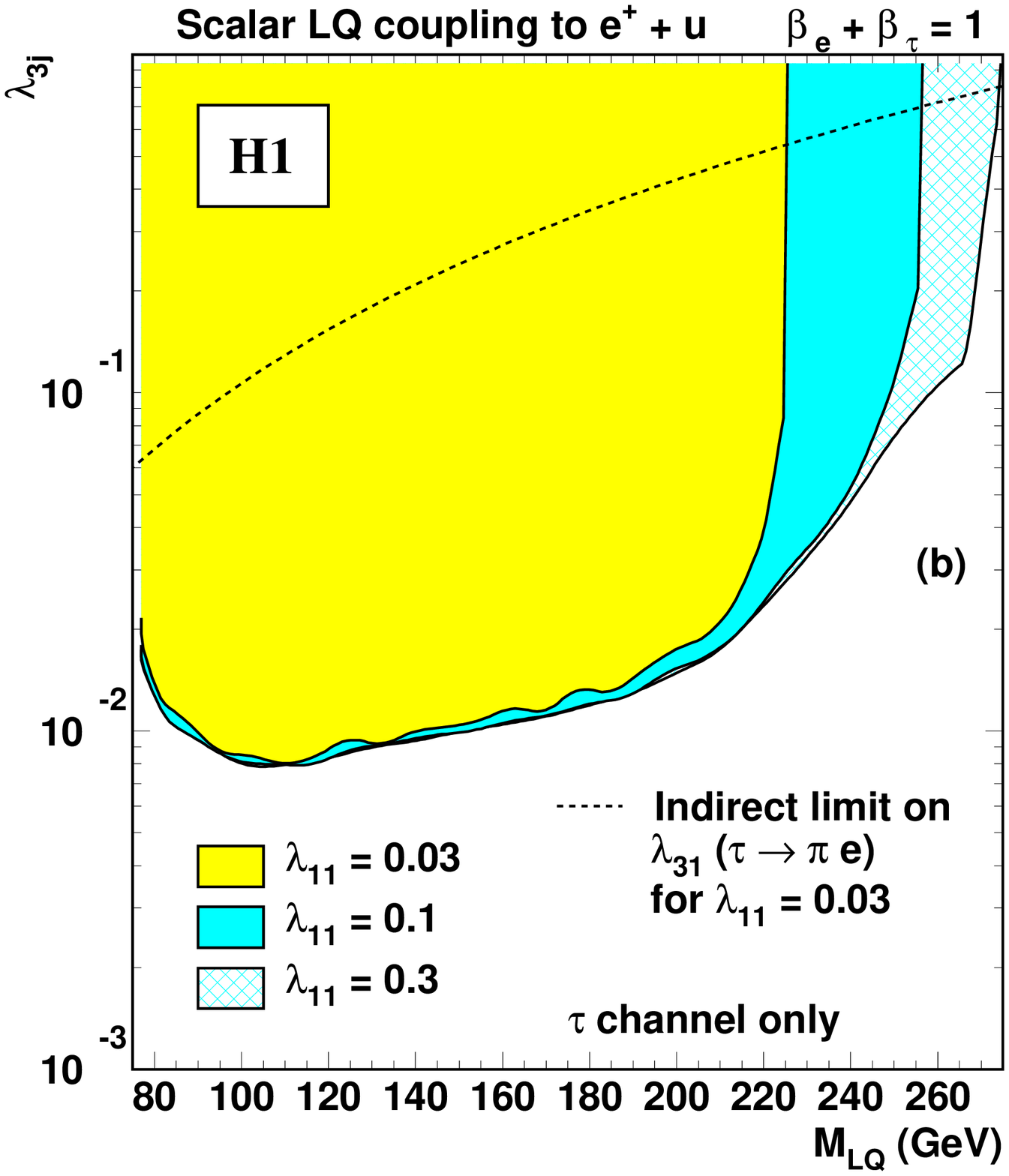,height=5.cm,width=5.5cm}
\begin{minipage}[t]{5.2cm}
{\fs Figure 3: Mass-dependent H1 limits on $\lambda_{\tau q_2}$ for a
scalar LQ coupling to $e^++u$, for different values of $\lambda_{e
q_1}$}
\end{minipage}
\end{minipage}\\

\noindent Figure 3 \cite{h1_lfv} shows mass-dependent limits on
$\lambda_{\tau q_2}$ for $S_{1/2}^L$ LQs for several values of $\lambda_{e
q_1}$. As can be seen, HERA has a substantial sensitivity on
$\lambda_{\tau q_2}$ for LQs which are light enough to be resonantly
produced via $\lambda_{e q_1}$. These LQs are assumed to couple only
to $eq$ and $\tau q$ (${\mathrm BR}_{e q_1}+{\mathrm BR}_{\tau
q_2}=1$). If both $\lambda_{e q_1}$ and $\lambda_{\tau q_2}$ are of
electromagnetic strength, HERA excludes such scalar LQs up to
270\,GeV. The TeVatron collider experiments complementarily exclude
third generation LQs (coupling to third generation fermions only) up to 99\,GeV
\cite{cdf_tau} (94\,GeV \cite{d0_tau}) for BR$_{\tau b}=1$
(${\mathrm BR}_{\nu b}=1$).\\
For all F=0 LQs with masses $M_{LQ}\gg\sqrt{s}$, limits on $\Psi_{q_1
q_2}^{\tau}$ are shown in Table~\ref{f0_tau_tab}. The HERA limits are
compared to the most stringent indirect limits \cite{h1_lfv}. The
superscript on the HERA limits indicates whether the strongest HERA
limit comes from ZEUS (Z) or from H1 (H) \cite{h1_lfv}. Although ZEUS
has a higher integrated luminosity and considers both the leptonic and
hadronic decay modes of the $\tau$, there are several cases where H1 reports
stronger limits. This is due to the fact that H1 assumes a common
selection efficiency for all quark combinations whereas ZEUS evaluates
the efficiencies for each possible $q_i q_j$-combination
individually. The ZEUS and H1 limits are stronger (bold numbers in
Tab. \ref{f0_tau_tab}) than those from low-energy measurements in
several slots, especially if higher generation quarks are
involved. ZEUS and H1 also set limits for some hitherto unconstrained
flavour combinations.   
   
\begin{table}[htb]
\begin{tabular}{|c || c | c | c | c || c | c | c |}\hline
$q_i q_j$ & $S_{1/2}^L$ & $S_{1/2}^R$ & $\tilde{S}_{1/2}^L$ & $V_0^L$     & $V_0^R$    & $\tilde{V}_0^R$& $V_1^L$ \\ \hline \hline
\fs 1 1 & \fs 0.030$^Z$  & \fs 0.025$^Z$  & \fs 0.046$^Z$  & \fs 0.033$^Z$ & \fs 0.033$^Z$  & \fs 0.024$^Z$  & \fs 0.012$^Z$ \\[-1mm]
     & \fs 0.0032$^a$ &  \fs 0.0016$^a$ &  \fs 0.0032$^a$ & \fs 0.002$^b$ & \fs 0.0016$^a$ & \fs 0.0016$^a$ & \fs 0.002$^b$ \\ \hline
\fs 1 2 & \fs \bf 0.030$^Z$  & \fs \bf 0.025$^Z$  & \fs \bf 0.046$^Z$ & \fs 0.036$^Z$ & \fs 0.036$^Z$  & \fs \bf 0.026$^Z$  & \fs 0.012$^Z$ \\[-1mm] 
        &                & \fs 0.05$^c$   & \fs 0.05$^c$   & \fs 0.03$^c$  & \fs 0.03$^c$   &                & \fs $2.5\cdot 10^{-6}$ $^d$ \\ \hline
\fs 1 3 &                & \fs \bf 0.049$^Z$  & \fs \bf 0.049$^Z$  & \fs 0.044$^Z$ & \fs 0.044$^Z$  &                & \fs 0.044$^Z$ \\[-1mm]
        &   *            & \fs 0.08$^e$   & \fs 0.08$^e$   & \fs 0.02$^f$  & \fs 0.04$^e$   &      *           & \fs 0.02$^f$  \\ \hline
\fs 2 1 & \fs \bf 0.15$^H$   & \fs 0.092$^Z$  & \fs 0.105$^Z$  & \fs 0.049$^Z$ & \fs 0.049$^Z$  & \fs \bf 0.061$^Z$  & \fs 0.026$^Z$ \\[-1mm]
         &                & \fs 0.05$^c$   & \fs 0.05$^c$   & \fs 0.03$^c$  & \fs 0.03$^c$   &                & \fs $2.5\cdot 10^{-6}$ $^d$ \\ \hline
\fs 2 2 & \fs 0.18$^H$   & \fs 0.10$^H$   & \fs \bf 0.120$^Z$  & \fs \bf 0.061$^Z$ & \fs \bf 0.061$^Z$  & \fs \bf 0.102$^Z$  & \fs \bf 0.041$^Z$ \\[-1mm]
        & \fs 0.03$^g$   & \fs 0.02$^g$   &                &               &                &                &       \\ \hline
\fs 2 3 &                & \fs 0.14$^H$   & \fs 0.14$^H$   & \fs 0.102$^Z$ & \fs 0.102$^Z$  &                & \fs 0.102$^Z$ \\[-1mm] 
        &   *            & \fs 0.08$^e$   & \fs 0.08$^e$   & \fs 0.02$^f$  & \fs 0.04$^e$   &  *             & \fs 0.02$^f$  \\ \hline
\fs 3 1 &                & \fs 0.16$^H$   & \fs 0.16$^H$   & \fs 0.052$^Z$ & \fs 0.052$^Z$  &                & \fs 0.052$^Z$ \\[-1mm] 
        &   *            & \fs 0.08$^e$   & \fs 0.08$^e$   & \fs 0.002$^h$ & \fs 0.04$^e$   &  *             & \fs 0.002$^h$  \\ \hline
\fs 3 2 &                & \fs 0.19$^H$   & \fs 0.19$^H$   & \fs 0.073$^Z$ & \fs 0.073$^Z$  &                & \fs 0.073$^Z$ \\[-1mm] 
    &      *             & \fs 0.08$^e$   &\fs  0.08$^e$   & \fs 0.02$^f$  & \fs 0.04$^e$   &  *             & \fs 0.02$^f$  \\ \hline
\fs 3 3 &                & \fs \bf 0.23$^H$   & \fs \bf 0.23$^H$   & \fs \bf 0.14$^H$  & \fs \bf 0.14$^H$   &                & \fs \bf 0.14$^H$  \\[-1mm] 
        &  *             &                &                & \fs0.51$^g$   & \fs 0.51$^g$   &  *             &  *     \\ \hline
\end{tabular}
\caption{95 $\%$ CL limits on $\Psi_{q_1 q_2}^{\tau}$ (in
$10^{-4}$\,GeV$^{-2}$) for F=0 leptoquarks. The first line in each row
shows the best HERA limits; the superscripts Z and H indicate ZEUS
or H1 results, respectively. The numbers in bold characters indicate
the cases where HERA provides the strongest limits. The second line in
each row states the most stringent low-energy constraint. The
superscripts indicate the respective low-energy measurement: (a) $\tau
\to \pi e$, (b) $G_F$, (c) $\tau \to K e$, (d) $K\to \pi \nu
\bar{\nu}$, (e) $B\to \tau e X$, (f) $B \to l \nu X$, (g) $\tau \to e
\gamma$, (h) $V_{ub}$. The * indicates the cases which would involve a
top quark. }\label{f0_tau_tab}  
\end{table} 
\section{Conclusions}
No evidence for LFV was found in the ZEUS and H1 1994-1997 
$e^+p$ data. Exclusion limits on $\lambda_{eq_1}\times
\sqrt{{\mathrm BR}_{\ell q_2}}$, $\lambda_{\tau q_2}$ and on
$\frac{\lambda_{e q_i}\lambda_{\ell q_j}}{M_{LQ}^2}$ have been set. Assuming
electromagnetic coupling strength, resonantly produced LFV LQs with
masses up to 260--285\,GeV are excluded. For LFV LQs with
$M_{LQ}\gg\sqrt{s}$, H1 and ZEUS set limits for hitherto unconstrained flavour
combinations and improve several existing limits, especially for
$e\leftrightarrow\tau$ transitions.   



\vspace*{-0.3cm}\section*{References}
\begin{thebibliography}{99}

\bibitem{sk}Y. Fukuda et al., Super-Kamiokande Collaboration, \Journal{\PRL}{81}{1562}{1998}

\bibitem{high_pt}C. Adloff et al., H1 Collaboration, \Journal{\EPJC}{5}{575}{1998}

\bibitem{h1_lfv}C. Adloff et al., H1 Collaboration, \Journal{\EPJC}{11}{447}{1999}

\bibitem{zeus_lfv}ZEUS Collaboration, Abstract 551, EPS99, Tampere,
July 1999

\bibitem{d0_mu}B. Abbott et al., D0 Collaboration, \Journal{\PRL}{84}{2088}{2000}

\bibitem{cdf_tau}F. Abe et al., CDF Collaboration, \Journal{\PRL}{78}{2906}{1997}

\bibitem{d0_tau}B. Abbott et al., D0 Collaboration, \Journal{\PRL}{81}{38}{1998}

\end{thebibliography}
\end{document}